\documentclass[11pt]{article}

\usepackage[preprint]{acl}

\usepackage{times}
\usepackage{latexsym}

\usepackage[T1]{fontenc}

\usepackage[utf8]{inputenc}

\usepackage{microtype}

\usepackage{inconsolata}

\usepackage{graphicx}

\usepackage{amsmath}
\usepackage{amsfonts}
\usepackage{booktabs,multirow,makecell}
\usepackage[table]{xcolor}
\usepackage{subcaption}
\usepackage{algorithm}
\usepackage{algpseudocode}
\usepackage{amsmath, amssymb}

\definecolor{ExcellentColor}{RGB}{220,245,220} 
\definecolor{GoodColor}{RGB}{235,245,255}      
\definecolor{FairColor}{RGB}{255,245,220}      
\definecolor{PoorColor}{RGB}{255,230,230}      
\definecolor{HeaderColor}{RGB}{80,120,160}

\title{Aligning Multimodal Sequential Recommendations via Robust Direct Preference Optimization with Sparse MoE}

\author{
  {\bf Hejin Huang}\textsuperscript{\rm 1},
  {\bf Jusheng Zhang}\textsuperscript{\rm 1},
  {\bf Kaitong Cai}\textsuperscript{\rm 1},
  {\bf Jian Wang}\textsuperscript{\rm 2},
  {\bf Rong Pan}\textsuperscript{\rm 1}\thanks{\ Corresponding author.}\\
   \textsuperscript{\rm 1}Sun Yat-sen University  
   \textsuperscript{\rm 2}Snap Inc \\
   \texttt{huanghj96@mail2.sysu.edu.cn} \\
}

\begin{document}
\maketitle
\begin{abstract} Preference-based alignment objectives have been widely adopted, from RLHF-style pairwise learning in large language models to emerging applications in recommender systems.
Yet, existing work rarely examines how Direct Preference Optimization (DPO) behaves under implicit feedback, where unobserved items are not reliable negatives.
We conduct systematic experiments on multimodal sequential recommendation to compare common negative-selection strategies and their interaction with DPO training.
Our central finding is that a simple modification---\textbf{replacing deterministic hard negatives with stochastic sampling from a dynamic top-$K$ candidate pool}---consistently improves ranking performance.
We attribute its effectiveness to two factors: (1) reducing erroneous suppressive gradients caused by false negatives, and (2) retaining informative hard signals while smoothing optimization via controlled stochasticity.
With an optional sparse Mixture-of-Experts encoder for efficient capacity scaling, RoDPO achieves up to \textbf{+5.25\%} NDCG@5 on three Amazon benchmarks, with nearly unchanged inference cost.
 \end{abstract}

\begin{figure*}[t]
  \centering
  \includegraphics[width=\textwidth]{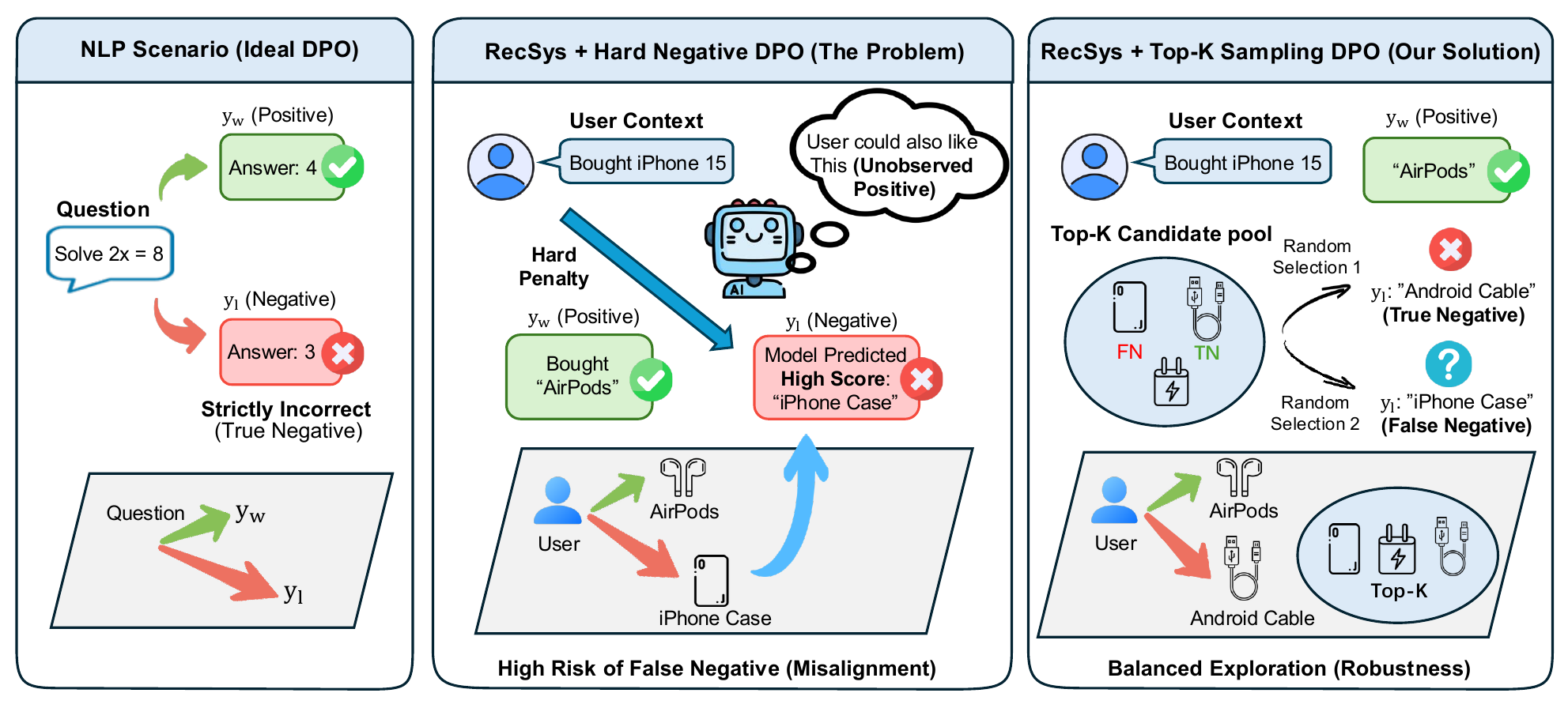}
 \caption{The ``False Negative'' Dilemma in RecSys DPO.
Left: In NLP, negatives are strictly incorrect, making DPO effective.
Middle: In RecSys, hard negatives may include unobserved positives, causing mis-penalization and embedding drift.
Right: Top-$K$ sampling adds randomness among candidates, balancing positives.}

  \label{fig:overview}
\end{figure*}

\section{Introduction}

Multimodal Sequential Recommendation (MSR) has become a pivotal component of modern personalized systems, leveraging rich side information (e.g., product images and textual descriptions) to model evolving user interests, especially under sparse interaction signals~\cite{zhang2019feature, zhou2020s3}. 
Most MSR methods follow the \textit{next-item prediction} paradigm and are trained with pointwise objectives such as Cross-Entropy (CE)~\cite{SASRec}. 
While effective, pointwise training primarily optimizes likelihood estimation and does not explicitly enforce \emph{relative} preference rankings: it implicitly treats all non-target items as equally negative, thereby overlooking nuanced ``better-than'' relations that are essential for decoding complex user intent, often leading to suboptimal ranking boundaries.

In parallel, preference optimization has reshaped alignment in NLP. 
Reinforcement Learning from Human Feedback (RLHF)~\cite{ouyang2022training} and Direct Preference Optimization (DPO)~\cite{rafailov2023direct} demonstrate that directly optimizing pairwise preference objectives can align models with human intent more effectively than likelihood-based training.
This success naturally raises a question for recommendation: \textit{Can we adapt DPO to align multimodal sequential recommenders with users' implicit preferences?}
A naïve transfer, however, breaks in recommender systems due to a domain-specific challenge: the \textit{false negative} dilemma, as illustrated in Figure~\ref{fig:overview}.
In NLP (Figure~\ref{fig:overview}, Left), negative samples are typically incorrect model-generated responses, which are strictly undesirable. 
In implicit-feedback recommendation, by contrast, an unclicked item is not necessarily disliked---it may simply be unobserved. 
Consequently, standard DPO practices such as mining \textit{hard negatives} (Figure~\ref{fig:overview}, Middle) can mistakenly choose unobserved positives and then \emph{aggressively suppress} them. 
This introduces erroneous preference gradients, distorts ranking space, and leads to severe performance degradation.

Importantly, existing attempts to import preference optimization into recommendation are often confounded by multiple design choices---backbone capacity, auxiliary objectives, and training schedules---obscuring the critical factors that govern DPO's efficacy under implicit feedback. 
This motivates a careful disentanglement: \textbf{Do performance gains stem from architectural complexity, or from rectifying DPO's negative sampling mechanism amid false negatives?}

In this work, we answer this question by systematically examining negative-selection strategies for DPO in MSR, including random negatives, argmax hard negatives, and stochastic sampling from a top-$K$ candidate pool. 
\textbf{Our central finding is that the effectiveness of DPO in recommendation hinges critically on the negative selection strategy.}
Specifically, we identify a simple modification---\textbf{Stochastic Top-$K$ Negative Sampling} (Figure~\ref{fig:overview}, Right)---as the optimal strategy.
Instead of deterministically penalizing the single argmax negative, we sample negatives from a dynamic top-$K$ pool of the model's highest-scoring candidates.
This approach retains the gradient intensity of hard negatives while introducing controlled stochasticity to substantially reduce the risk of colliding with false negatives.

Based on this finding, we propose \textbf{RoDPO} (\textbf{Ro}bust \textbf{D}irect \textbf{P}reference \textbf{O}ptimization), a framework tailored for multimodal sequential recommendation. 
RoDPO is plug-and-play: it preserves the DPO objective and backbone architecture, while replacing only the negative-selection rule to handle the implicit-feedback regime.
To further ensure sufficient capacity for resolving fine-grained preference conflicts, we optionally incorporate a \textbf{sparse Mixture-of-Experts (MoE)} encoder~\cite{shazeer2017outrageously} for efficient scaling, and adopt a \textbf{two-stage warm-up} protocol to stabilize optimization by anchoring the reference policy on a reliable latent space. 

Our contributions are summarized as follows:
\begin{itemize}
\item We identify \textit{false negatives} as the key source of DPO brittleness under implicit feedback, and show that negative selection is critical.
    \item We propose \textbf{Stochastic Top-$K$ Negative Sampling}, a simple plug-and-play modification to DPO that balances hardness and false-negative tolerance.
    \item Experiments on three Amazon benchmarks show consistent gains over strong multimodal baselines, achieving up to \textbf{+5.25\%} NDCG@5 with negligible inference overhead.
\end{itemize}

\section{Related Work}
\subsection{Multimodal Sequential Recommendation}
Sequential Recommendation (SR) predicts future interactions based on historical behaviors. While early methods like SASRec~\cite{SASRec} and BERT4Rec~\cite{sun2019bert4rec} relied on ID embeddings, Multimodal SR (MSR) incorporates side information to alleviate sparsity. Representative works focus on sophisticated fusion mechanisms: FDSA~\cite{zhang2019feature} uses feature-level self-attention, NOVA~\cite{NOVA} proposes non-invasive fusion, and HM4SR~\cite{HM4SR} employs heuristic alignment tasks. However, these methods primarily rely on pointwise (e.g., Cross-Entropy) or simple pairwise losses (e.g., BPR~\cite{rendle2009bpr}). They prioritize \textit{feature representation} over the explicit alignment of the output distribution with user preference rankings, which is the core focus of our DPO-based framework.

\subsection{Preference Alignment via DPO}
Aligning generative models with human intent has evolved from Reinforcement Learning from Human Feedback (RLHF)~\cite{ouyang2022training} to Direct Preference Optimization (DPO)~\cite{rafailov2023direct}, which derives a closed-form objective to optimize policies without an explicit reward model. While DPO is standard in NLP, its application in recommendation is nascent. Recent attempts like DPO4Rec~\cite{sun2025direct} explore DPO for explanation generation or re-ranking. In contrast, we adapt DPO for the core ranking task in multimodal sequential recommendation, specifically designing a robust negative sampling strategy to mitigate the false negative dilemma inherent in implicit feedback.

\subsection{Negative Sampling in Recommendation}
Negative sampling is critical for implicit feedback. Standard approaches sample uniformly from non-interacted items~\cite{rendle2009bpr}, while Hard Negative Mining (e.g., DNS~\cite{zhang2013optimizing}, MixGCF~\cite{huang2021mixgcf}) accelerates convergence by selecting indistinguishable items. However, we argue traditional hard mining is perilous for DPO. Unlike metric learning where hard negatives define decision boundaries, DPO actively suppresses the ``losing'' item $y_l$. In sparse datasets, ``hard'' items are often \textit{False Negatives} (unobserved positives)~\cite{he2017neural}. Penalizing them introduces erroneous gradients. Our work identifies this conflict and proposes Stochastic Top-K Sampling to reconcile gradient intensity with false-negative robustness.

\begin{figure*}[t]
  \centering
  \includegraphics[width=\textwidth]{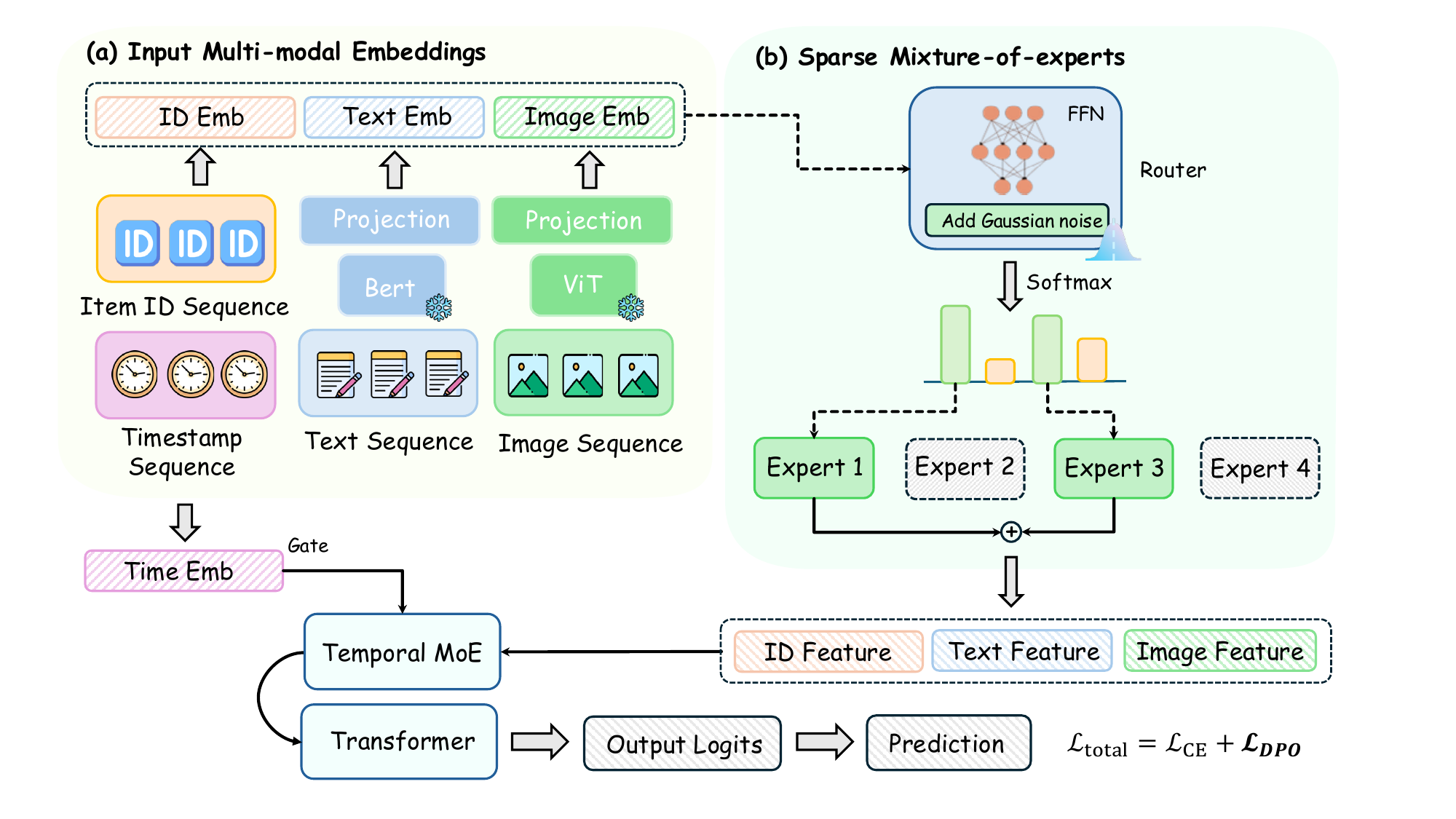}
  \caption{Overall framework of \textbf{RoDPO}. 
  (a) \textbf{Multimodal Encoder}: Item IDs and text/image features are embedded into a shared space, refined by a sparse MoE, and modeled by modality-specific Transformers with a temporal MoE for next-item prediction. 
  (b) \textbf{Robust DPO}: Preference pairs are formed by the observed item $y_w$ and a sampled negative $y_l$ from a dynamic top-$K$ pool (excluding $y_w$), optimized against a frozen reference policy.}
  \label{fig:framework}
\end{figure*}

\section{Methodology}
\label{sec:method}

In this section, we elaborate on the proposed \textbf{RoDPO} framework.
We first formalize Multimodal Sequential Recommendation (MSR) in \S\ref{sec:problem} and describe our multimodal sequential encoder in \S\ref{sec:encoder} (Fig.~\ref{fig:framework}a).
We then analyze why naïve DPO is brittle under implicit feedback in \S\ref{sec:analysis}.
Based on this analysis, we introduce our core contribution---\textbf{Stochastic Top-$K$ Negative Sampling}---in \S\ref{sec:rodpo} (Fig.~\ref{fig:framework}b).
Finally, we present the overall training objective and a robust two-stage protocol.
The overall workflow is summarized in Algorithm~\ref{alg:rodpo}.

\subsection{Problem Setup}
\label{sec:problem}
In MSR, a training instance consists of a user interaction sequence $x$ augmented with multimodal attributes:
\[
x = \{(i_1, m_1, t_1), \ldots, (i_L, m_L, t_L)\},
\]
where $i_\ell \in \mathcal{I}$ is the item ID from the universe $\mathcal{I}$, $m_\ell$ represents multimodal attributes (e.g., visual and textual features), and $t_\ell$ denotes the timestamp.
The goal is to predict the next item $y^+ \in \mathcal{I}$ given context $x$.
The model parameterizes a scoring function (logit) for ranking:
\begin{equation}
s_\theta(x, i) \in \mathbb{R}, \quad \forall i \in \mathcal{I}.
\label{eq:score}
\end{equation}

\subsection{Multimodal Sequential Encoder}
\label{sec:encoder}
As illustrated in Figure~\ref{fig:framework}(a), RoDPO encodes item IDs and multimodal signals with modality-specific Transformers, optionally refined by a sparse MoE layer and a temporal module.
Our framework is backbone-agnostic; without loss of generality, we adopt a Transformer-based architecture for sequential modeling.
\textbf{Multimodal Embeddings.}
We project discrete IDs and pre-extracted multimodal features into a shared $d$-dimensional space.
Specifically, $\mathbf{e}^{\text{id}}_\ell=\mathrm{Emb}(i_\ell)$,
$\mathbf{e}^{\text{txt}}_\ell=\mathbf{W}_{\text{txt}}\mathbf{v}^{\text{txt}}_\ell$, and
$\mathbf{e}^{\text{img}}_\ell=\mathbf{W}_{\text{img}}\mathbf{v}^{\text{img}}_\ell$,
where $\mathbf{v}^{\text{txt}}_\ell$ and $\mathbf{v}^{\text{img}}_\ell$ are pre-extracted text/image features and
$\mathbf{W}_{\text{txt}}, \mathbf{W}_{\text{img}}$ are learnable projections.
Absolute positional embeddings are added to preserve sequential order.

\paragraph{(Optional) Sparse Noisy MoE Refinement.}
To optionally scale model capacity for multimodal fusion with limited inference overhead, we incorporate a \textbf{Sparse Noisy Mixture-of-Experts (MoE)} layer (Fig.~\ref{fig:framework}a).
Given an input representation $h$, a router selects top-$k$ experts and aggregates their outputs:
\begin{equation}
\text{MoE}(h) = \sum_{j \in \mathcal{S}(h)} G(h)_j \cdot E_j(h), \quad |\mathcal{S}(h)| = k,
\label{eq:moe}
\end{equation}

where $E_j(\cdot)$ denotes the $j$-th expert network (MLP), $\mathcal{S}(h)$ is the selected expert set, and $G(h)$ is a sparse gating function.
Following common practice, we inject Gaussian noise into gating logits during training to improve load balancing.
This component is orthogonal to our core contribution in \S\ref{sec:rodpo} and can be enabled when additional capacity is desired.

\paragraph{Temporal Fusion and Logits.}
We obtain modality-specific sequence representations via modality-specific Transformer encoders and a timestamp-conditioned temporal module (Fig.~\ref{fig:framework}a).
The final score for candidate item $i$ is computed by a weighted fusion:
\begin{equation}
s_\theta(x,i) = s^{\text{id}}_\theta(x,i) + \alpha_{\text{txt}}\, s^{\text{txt}}_\theta(x,i) + \alpha_{\text{img}}\, s^{\text{img}}_\theta(x,i),
\label{eq:fusion}
\end{equation}

where $\alpha_{\text{txt}}$ and $\alpha_{\text{img}}$ are learnable scalars.

\subsection{The Challenge: Naïve DPO and False Negatives}
\label{sec:analysis}
Direct Preference Optimization (DPO) aligns a policy $\pi_\theta$ with a reference policy $\pi_{\text{ref}}$ by optimizing pairwise preferences.
While effective in NLP, standard DPO strategies become brittle in recommendation due to the \textit{False Negative} dilemma under implicit feedback.
In this setting, an unobserved item is not necessarily disliked; it may simply be unexposed.
Consequently, \textbf{hard negative mining} (e.g., selecting the argmax non-target item) can frequently select \emph{unobserved positives}.
Penalizing such false negatives introduces erroneous preference gradients, causing overly suppressive updates and distorting the learned ranking space.

\subsection{RoDPO: Robust DPO with Stochastic Top-$K$ Negatives}
\label{sec:rodpo}
Figure~\ref{fig:framework}(b) summarizes RoDPO's preference construction and robust DPO alignment.

\paragraph{Preference Construction.}
For a context $x$, the ground-truth next item is treated as the positive (winning) item $y_w = y^+$.
The key challenge is to construct a losing item $y_l$ that is informative yet robust to false negatives.

\paragraph{Stochastic Top-$K$ Negative Sampling (Core Fix).}
Instead of deterministically setting $y_l$ as the single hardest negative, we construct a dynamic candidate pool from the model's top-$K$ predictions excluding $y_w$:
\begin{equation}
\mathcal{C}_K(x) = \operatorname{TopK}\left( \{ s_\theta(x, i) \mid i \in \mathcal{I} \setminus \{y_w\} \}, K \right).
\label{eq:ck}
\end{equation}

We then sample the losing item uniformly:
$
y_l \sim \mathcal{U}(\mathcal{C}_K(x)).$This minimal modification preserves informative ``hard'' signals (since $\mathcal{C}_K(x)$ contains high-scoring candidates) while reducing repeated collisions with a specific false negative, thereby stabilizing preference optimization under implicit feedback.
\paragraph{Mechanism, Objective, and Optimization.}
Stochastic Top-$K$ sampling induces a principled trade-off:
\textbf{(i) Informative hardness}---restricting candidates to the top-$K$ maintains non-vanishing gradients and avoids the inefficiency of purely random negatives; and
\textbf{(ii) False-negative tolerance}---stochasticity reduces overly suppressive updates caused by repeatedly penalizing the same unobserved positive, smoothing optimization and improving robustness.
Moreover, RoDPO only requires the logits of $(y_w, y_l)$ to form the preference margin, since the softmax partition function cancels in the log-ratio.
We therefore write DPO directly with logit margins:
\begin{equation}
\begin{aligned}
\mathcal{L}_{\text{DPO}}(x, y_w, y_l)
= -\log \sigma \Big( \beta \big[ \Delta s_\theta(x, y_w, y_l) \\
- \Delta s_{\text{ref}}(x, y_w, y_l) \big] \Big),
\label{eq:dpo}
\end{aligned}
\end{equation}

where $\Delta s(x, y_w, y_l) = s(x, y_w) - s(x, y_l)$ and $\beta$ controls the strength of preference enforcement relative to the frozen reference policy.
In practice, we adopt a simple two-stage optimization: we first warm up the backbone with standard CE next-item prediction to obtain $\pi_{\text{sft}}$, then set the frozen reference $\pi_{\text{ref}} \leftarrow \pi_{\text{sft}}$ and optimize the joint objective:
\begin{equation}
\mathcal{L}_{\text{total}} = \mathcal{L}_{\text{CE}} + \lambda\, \mathcal{L}_{\text{DPO}}.
\label{eq:total}
\end{equation}

Algorithm~\ref{alg:rodpo} summarizes the Stage-2 RoDPO optimization with stochastic Top-$K$ negatives.

\begin{algorithm}[t]
\caption{RoDPO (Stage~2) Training with Stochastic Top-$K$ Sampling}
\label{alg:rodpo}
\begin{algorithmic}[1]
\Require Training dataset $\mathcal{D}$, item set $\mathcal{I}$, hyperparameters $K, \beta, \lambda$.
\Require Warmed-up model $\pi_{\text{sft}}$ from Stage~1.
\State \textbf{Initialize:} policy $\pi_\theta \leftarrow \pi_{\text{sft}}$, reference $\pi_{\text{ref}} \leftarrow \pi_{\text{sft}}$.
\State \textbf{Freeze:} parameters of $\pi_{\text{ref}}$.
\While{not converged}
    \For{each batch $B=\{(x, y_w)\}$ sampled from $\mathcal{D}$}
        \State \textbf{// 1. Forward pass (Fig.~\ref{fig:framework}a)}
        \State Compute logits $s_\theta(x,\cdot)$ using the encoder in Fig.~\ref{fig:framework}(a) and fusion in Eq.~\eqref{eq:fusion}.
        \State Compute reference logits $s_{\text{ref}}(x,\cdot)$ with $\pi_{\text{ref}}$ (no grad).
        \State \textbf{// 2. Stochastic Top-$K$ negative sampling (Fig.~\ref{fig:framework}b)}
        \State $\mathcal{C}_K(x) \leftarrow \operatorname{TopK}\!\left(\{s_\theta(x,i)\mid i\neq y_w\}, K\right)$.
        \State $y_l \sim \mathcal{U}(\mathcal{C}_K(x))$.
        \State \textbf{// 3. Compute losses}
        \State $\Delta s_\theta \leftarrow s_\theta(x,y_w) - s_\theta(x,y_l)$; \ $\Delta s_{\text{ref}} \leftarrow s_{\text{ref}}(x,y_w) - s_{\text{ref}}(x,y_l)$.
        \State $\mathcal{L}_{\text{DPO}} \leftarrow -\log \sigma(\beta (\Delta s_\theta - \Delta s_{\text{ref}}))$ \Comment{Eq.~\eqref{eq:dpo}}
        \State $\mathcal{L}_{\text{CE}} \leftarrow \text{CrossEntropy}(s_\theta(x,\cdot), y_w)$.
        \State \textbf{// 4. Update}
        \State $\mathcal{L}_{\text{total}} \leftarrow \mathcal{L}_{\text{CE}} + \lambda \mathcal{L}_{\text{DPO}}$ \Comment{Eq.~\eqref{eq:total}}
        \State Update $\theta$ via gradient descent on $\mathcal{L}_{\text{total}}$.
    \EndFor
\EndWhile
\end{algorithmic}
\end{algorithm}

\section{Experiments}

In this section, we conduct extensive experiments to evaluate the effectiveness of RoDPO. We aim to answer the following research questions:
\textbf{RQ1}: Does RoDPO outperform state-of-the-art multimodal sequential recommendation methods?
\textbf{RQ2}: How do different negative sampling strategies (e.g., Random vs. Hard vs. Top-K) impact the preference alignment process?
 \textbf{RQ3}: What is the contribution of each component (e.g., Sparse MoE, DPO) to the final performance?

\begin{table*}[t]
\centering
\renewcommand{\arraystretch}{1.2}
\vspace{-10pt}
\resizebox{\textwidth}{!}{
\begin{tabular}{l|cccc|cccc|cccc}
\toprule
\multicolumn{1}{l|}{\multirow{3}{*}{\textbf{Method}}} &
  \multicolumn{4}{c|}{\textbf{Toys and Games}} &
  \multicolumn{4}{c|}{\textbf{Beauty}} &
  \multicolumn{4}{c}{\textbf{Home and Kitchen}} \\ 
  \cmidrule(lr){2-5} \cmidrule(lr){6-9} \cmidrule(lr){10-13}
\multicolumn{1}{c|}{} &
  \multicolumn{2}{c}{NDCG} &
  \multicolumn{2}{c|}{MRR} &
  \multicolumn{2}{c}{NDCG} &
  \multicolumn{2}{c|}{MRR} &
  \multicolumn{2}{c}{NDCG} &
  \multicolumn{2}{c}{MRR} \\
\multicolumn{1}{c|}{} &
  @5 &
  @10 &
  @5 &
  @10 &
  \multicolumn{1}{c}{@5} &
  \multicolumn{1}{c}{@10} &
  \multicolumn{1}{c}{@5} &
  \multicolumn{1}{c|}{@10} &
  \multicolumn{1}{c}{@5} &
  \multicolumn{1}{c}{@10} &
  \multicolumn{1}{c}{@5} &
  \multicolumn{1}{c}{@10} \\ \midrule
  \rowcolor{HeaderColor}
  \multicolumn{13}{c}{\textcolor{white}{\textbf{Traditional Non-time-aware Methods}}} \\
  \midrule
      
\rowcolor{PoorColor}
GRU4Rec &
  0.0236 &
  0.0289 &
  0.0201 &
  0.0222 &
  0.0271 &
  0.0337 &
  0.0223 &
  0.0260 &
  0.0067 &
  0.0087 &
  0.0056 &
  0.0064 \\
\rowcolor{PoorColor}
SASRec &
  0.0348 &
  0.0411 &
  0.0257 &
  0.0295 &
  0.0325 &
  0.0415 &
  0.0246 &
  0.0283 &
  0.0118 &
  0.0148 &
  0.0089 &
  0.0100 \\
\rowcolor{PoorColor}
LRURec &
  0.0362 &
  0.0446 &
  0.0278 &
  0.0312 &
  0.0323 &
  0.0412 &
  0.0250 &
  0.0286 &
  0.0112 &
  0.0141 &
  0.0084 &
  0.0096 \\ 
  \midrule
  \rowcolor{HeaderColor}
  \multicolumn{13}{c}{\textcolor{white}{\textbf{Traditional Time-aware Methods}}} \\
  \midrule
\rowcolor{FairColor}
TiSASRec &
  0.0372 &
  0.0467 &
  0.0278 &
  0.0317 &
  0.0340 &
  0.0432 &
  0.0264 &
  0.0302 &
  0.0117 &
  0.0146 &
  0.0088 &
  0.0100 \\
\rowcolor{FairColor}
FEARec &
  0.0354 &
  0.0442 &
  0.0266 &
  0.0302 &
  0.0342 &
  0.0434 &
  0.0265 &
  0.0304 &
  0.0121 &
  0.0152 &
  0.0092 &
  0.0105 \\
\rowcolor{FairColor}
TiCoSeRec &
  0.0336 &
  0.0404 &
  0.0280 &
  0.0311 &
  0.0335 &
  0.0413 &
  0.0285 &
  0.0317 &
  0.0108 &
  0.0134 &
  0.0091 &
  0.0102 \\ 
  \midrule
  \rowcolor{HeaderColor}
  \multicolumn{13}{c}{\textcolor{white}{\textbf{Multi-modal Methods}}} \\
  \midrule
\rowcolor{GoodColor}
NOVA &
  0.0381 &
  0.0478 &
  0.0288 &
  0.0328 &
  0.0347 &
  0.0442 &
  0.0273 &
  0.0312 &
  0.0120 &
  0.0147 &
  0.0092 &
  0.0102 \\
\rowcolor{GoodColor}
DIF-SR &
  0.0359 &
  0.0444 &
  0.0284 &
  0.0318 &
  0.0340 &
  0.0436 &
  0.0262 &
  0.0302 &
  0.0092 &
  0.0112 &
  0.0078 &
  0.0086 \\
\rowcolor{GoodColor}
UniSRec &
  0.0276 &
  0.0384 &
  0.0194 &
  0.0239 &
  0.0287 &
  0.0398 &
  0.0212 &
  0.0288 &
  0.0121 &
  0.0160 &
  0.0092 &
  0.0108 \\
\rowcolor{GoodColor}
MISSRec &
  0.0323 &
  0.0414 &
  0.0235 &
  0.0270 &
  0.0315 &
  0.0397 &
  0.0240 &
  0.0271 &
  0.0140 &
  0.0174 &
  0.0107 &
  0.0119 \\
\rowcolor{GoodColor}
M3SRec &
  0.0416 &
  0.0486 &
  0.0363 &
  0.0391 &
  0.0345 &
  0.0428 &
  0.0294 &
  0.0328 &
  0.0122 &
  0.0144 &
  0.0105 &
  0.0115 \\
\rowcolor{GoodColor}
IISAN &
  0.0395 &
  0.0489 &
  0.0354 &
  0.0393 &
  0.0372 &
  0.0464 &
  0.0309 &
  0.0347 &
  0.0152 &
  \underline{0.0187} &
  0.0129 &
  0.0142 \\
\rowcolor{GoodColor}
TedRec &
  0.0318 &
  0.0397 &
  0.0265 &
  0.0297 &
  0.0330 &
  0.0419 &
  0.0275 &
  0.0311 &
  0.0113 &
  0.0140 &
  0.0094 &
  0.0105 \\
\rowcolor{GoodColor}
HM4SR$\dagger$ &
  \underline{0.0495} &
  \underline{0.0572} &
  \underline{0.0430} &
  \underline{0.0462} &
  \underline{0.0413} &
  \underline{0.0489} &
  \underline{0.0356} &
  \underline{0.0388} &
  \underline{0.0162} &
  \underline{0.0187} &
  \underline{0.0142} &
  \underline{0.0153} \\ 
  \midrule
  \rowcolor{HeaderColor}
  \multicolumn{13}{c}{\textcolor{white}{\textbf{Our Methods}}} \\
  \midrule
\rowcolor{ExcellentColor}
\textbf{RoDPO} &
  \textbf{0.0521} &
  \textbf{0.0589} &
  \textbf{0.0463} &
  \textbf{0.0491} &
  \textbf{0.0436} &
  \textbf{0.0508} &
  \textbf{0.0379} &
  \textbf{0.0409} &
  \textbf{0.0165} &
  \textbf{0.0191} &
  \textbf{0.0146} &
  \textbf{0.0157} \\
\rowcolor{ExcellentColor}
\textit{Improv.} &
  +5.25\% &
  +2.97\% &
  +7.67\% &
  +6.28\% &
  +5.57\% &
  +3.89\% &
  +6.46\% &
  +5.41\% &
  +1.85\% &
  +2.14\% &
  +2.82\% &
  +2.61\% \\ \bottomrule
\end{tabular}
}
\caption{Overall performance comparison. The best results are bolded, and the second best are
underlined. Improvements are calculated against the
best baseline. $\dagger$ denotes that the corresponding method is reproduced by ourselves.}
\label{tab:main-exp}
\end{table*}

\subsection{Experimental Settings}

\paragraph{Datasets.}
We evaluate our method on three widely used benchmarks from the Amazon review dataset \cite{he2016ups}: \textbf{Toys and Games}, \textbf{Beauty}, and \textbf{Home and Kitchen}. These datasets contain user interaction sequences with rich multimodal information (product images and descriptions). Following standard preprocessing protocols \cite{zhou2020s3}, we filter out users and items with fewer than 5 interactions and treat the last item in the sequence as the test target, the second-to-last as validation, and the rest for training. The statistics of the processed datasets are shown in Appendix~\ref{sec:dataset_statistics}.


\paragraph{Baselines.}
To evaluate the effectiveness of RoDPO, we compare it against competitive baselines from three categories:
\textbf{(1) Traditional Non-Time-Aware Models:} GRU4Rec~\cite{GRU4Rec}, SASRec~\cite{SASRec}, and LRURec~\cite{LRURec}.
\textbf{(2) Traditional Time-Aware Models:} TiSASRec~\cite{TiSASRec}, FEARec~\cite{FEARec}, and TiCoSeRec~\cite{TiCoSeRec}.
\textbf{(3) Multimodal Models:} NOVA~\cite{NOVA}, DIF-SR~\cite{DIF-SR}, UniSRec~\cite{UniSRec}, MISSRec~\cite{MISSRec}, M3SRec~\cite{M3SRec}, IISAN~\cite{IISAN}, TedRec~\cite{TedRec}, and the state-of-the-art HM4SR~\cite{HM4SR}.
Detailed descriptions of these baselines are provided in Appendix~\ref{sec:baselines}.


\paragraph{Implementation Details.}
We implement RoDPO using PyTorch, leveraging the widely adopted open-source library RecBole\footnote{https://github.com/RUCAIBox/RecBole}~\cite{Recbole}. The embedding dimension is set to 64, and the maximum sequence length is 50. The Sparse MoE layer consists of 4 experts with $K=2$ active experts.
We adopt a \textbf{Two-Stage Training} protocol: 
In the first stage, we warm up the model using $\mathcal{L}_{\text{SFT}}$ for 15 epochs to ensure a stable reference model. 
In the second stage, we fine-tune with the DPO objective using $\beta=1.0$ and a learning rate of 1e-3. 
For our proposed \textit{Stochastic Top-K Sampling}, we set $K=50$. 
Evaluation metrics include \textbf{NDCG@K} and \textbf{MRR@K} with $K=\{5, 10\}$. For more details, please refer to the Appendix
~\ref{sec:metrics_detail}.

\subsection{Overall Performance (RQ1)}

Table \ref{tab:main-exp} presents the performance comparison on three Amazon benchmarks. 
RoDPO consistently outperforms all baselines across all metrics. Specifically:
\begin{itemize}
    \item Compared to the strongest baseline HM4SR, RoDPO achieves a significant improvement of \textbf{5.25\%} in NDCG@5 and \textbf{7.67\%} in MRR@5 on the \textsc{Toys and Games} dataset. This indicates that explicitly aligning user preferences via DPO provides a substantial advantage over purely pointwise training objectives.
    \item The improvement is consistent across different cutoff values ($K$), demonstrating the robustness of our ranking capability.
\end{itemize}

\subsection{Analysis of Negative Sampling (RQ2)}

A core contribution of this work is the \textit{Stochastic Top-K Negative Sampling}. To validate its effectiveness, we compare it with two variants:
(1) \textbf{Random Sampling}: Uniformly sampling from non-interacted items.
(2) \textbf{Hard Sampling (Argmax)}: Selecting the non-target item with the highest predicted probability.

As shown in Table \ref{tab:sampling_ablation}, \textbf{Hard Sampling (Argmax)} only marginally outperforms \textbf{Random Sampling}. This limited gain confirms that aggressive hard mining can introduce ``false negatives'' in sparse datasets, leading to sub-optimal training. In contrast, our \textbf{Top-K Sampling} achieves the best results (e.g., \textbf{0.0521} in NDCG@5), significantly surpassing the argmax baseline. By sampling from a top-$K$ pool rather than a single point, our method strikes a superior balance: it provides strong gradient signals while mitigating false-negative risks through stochasticity.


\begin{table}[h]
\centering
\resizebox{\columnwidth}{!}{
\begin{tabular}{lcc}
\toprule
\textbf{Sampling Strategy} & \textbf{NDCG@5} & \textbf{MRR@5} \\
\midrule
Random Sampling & 0.0491 & 0.0437\\
Hard Sampling (Argmax) & 0.0497 & 0.0438\\
\textbf{Top-K Sampling (Ours)} & \textbf{0.0521} & \textbf{0.0463} \\
\bottomrule
\end{tabular}
}
\caption{Ablation study on negative sampling strategies.}
\label{tab:sampling_ablation}
\end{table}



\subsection{Ablation Study (RQ3)}

We further investigate the impact of the Sparse MoE module and the DPO objective itself. 
Table \ref{tab:module_ablation} shows that:
(1) Removing DPO (i.e., using only SFT) results in a noticeable performance drop, confirming the necessity of preference alignment.
(2) Replacing Sparse MoE with a standard MLP (Base + DPO) yields a score of 0.0516, which is lower than the full model (0.0521). This suggests that the increased capacity from Sparse MoE helps the model better fit the complex preference boundaries defined by DPO.

\begin{table}[h]
\centering
\resizebox{\columnwidth}{!}{
\begin{tabular}{lcc}
\toprule
\textbf{Variant} & \textbf{NDCG@5} & \textbf{MRR@5}\\
\midrule
Full Model (RoDPO) & \textbf{0.0521} & \textbf{0.0463}\\
w/o Sparse MoE & 0.0516 & 0.0459 \\
w/o DPO (Baseline) & 0.0495 & 0.0430 \\
\bottomrule
\end{tabular}
}
\caption{Ablation on model components.}
\label{tab:module_ablation}
\end{table}

\section{Analysis}

In this section, we provide a comprehensive analysis to understand the robustness and efficiency of RoDPO. We focus on hyperparameter sensitivity, training efficiency, the visualization of learned preference distributions, and case study.

\subsection{Hyperparameter Sensitivity}

\begin{figure}[t] 
\centering 
\begin{subfigure}[t]{0.48\linewidth} 
\centering 
\includegraphics[width=\linewidth]{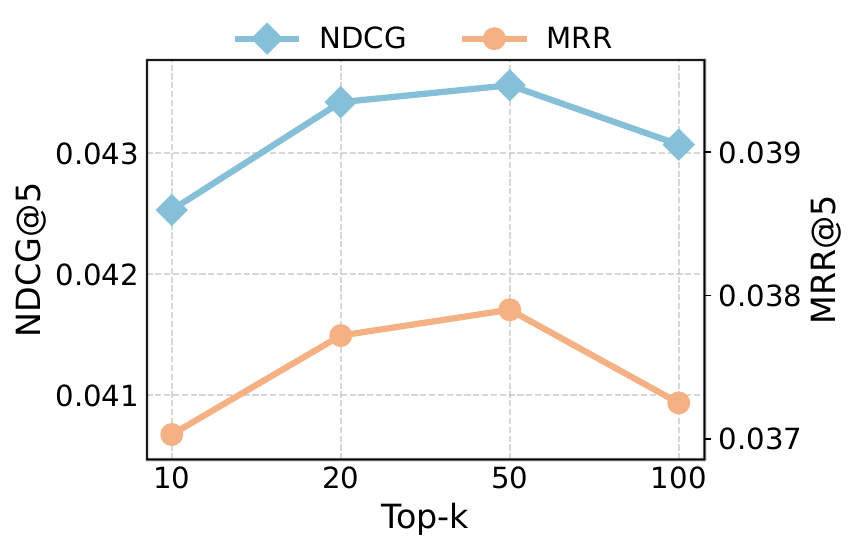}
\end{subfigure}
\begin{subfigure}[t]{0.48\linewidth} 
\centering 
\includegraphics[width=\linewidth]{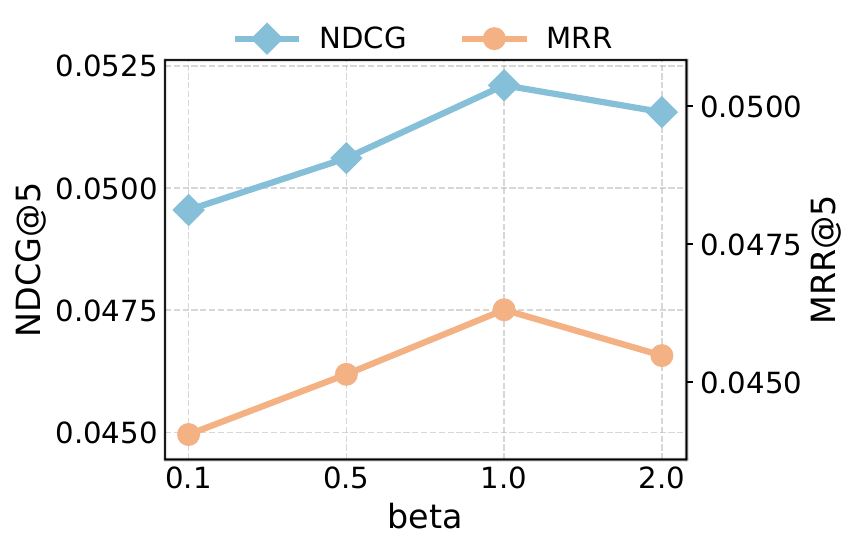}
\end{subfigure}
\caption{Effect of Top-$k$ Negative Sampling and DPO Coefficient $\beta$ on NDCG@5 and MRR@5.}
\label{fig:sensitivity_k_beta}
\end{figure}

\paragraph{Impact of Negative Sampling Size $K$.}
The parameter $K$ in our Stochastic Top-K Sampling strategy controls the trade-off between gradient intensity (hardness) and exploration (randomness). We vary $K$ in $\{10, 20, 50, 100\}$ on the Amazon \textsc{Toys and Games} dataset.
As illustrated in Figure \ref{fig:sensitivity_k_beta} (left), the performance initially improves as $K$ increases, peaking at $K=50$. 
\begin{itemize}
    \item When $K$ is small (e.g., $K=10$), the sampling approaches \textit{Hard Negative Sampling}, leading to overfitting on false negatives and a drop in NDCG.
    \item When $K$ is too large (approaching random sampling), the difficulty of negative samples decreases, resulting in vanishing gradients and slower convergence.
\end{itemize}
This observation validates our hypothesis that a moderate $K$ optimally balances the alignment objective.

\paragraph{Impact of DPO Coefficient $\beta$.}
The coefficient $\beta$ controls the strength of the KL-divergence constraint towards the reference model. We test $\beta \in \{0.1, 0.5, 1.0, 2.0\}$. 
Results show that $\beta=1.0$ yields the best performance. Unlike NLP tasks where a smaller $\beta$ (e.g., 0.1) is often preferred to prevent language degeneration, in RecSys, a larger $\beta$ forces the model to distinguish items more aggressively. Since our reference model is well-initialized (via the Two-Stage protocol), a stronger constraint ensures more stable alignment.

\subsection{Efficiency Analysis}

A common concern with DPO and MoE architectures is the potential increase in computational cost. We compare RoDPO with the strongest baseline HM4SR in terms of trainable model parameters, training time per epoch, and inference latency. The statistics are recorded on a single NVIDIA 4090 GPU.

\begin{table}[h]
\centering
\resizebox{\columnwidth}{!}{
\begin{tabular}{lcccc}
\toprule
\textbf{Model} & \makecell{\textbf{Trainable} \\ \textbf{Params}} & \textbf{Training} & \textbf{Inference} & \textbf{NDCG@5} \\
 & \textbf{(M)} & \textbf{(s/epoch)} & \textbf{(s/batch)} & \\
\midrule
HM4SR & 1.59 & 9.81 & 0.42 & 0.0495 \\
\textbf{RoDPO} & 1.60 & 9.84 & 0.45 & \textbf{0.0521} \\
\bottomrule
\end{tabular}
}
\caption{Efficiency comparison. ``Inference Latency'' denotes the time to process a batch of size 2048.}
\label{tab:efficiency}
\end{table}

Table \ref{tab:efficiency} reveals two key insights:
(1) \textbf{Comparable Training Cost}: Despite the DPO framework involving a reference model, \textbf{RoDPO} incurs negligible training overhead ($9.84$s vs. $9.81$s per epoch). This is because the reference model remains frozen and shares the same backbone architecture, ensuring the training remains highly efficient.
(2) \textbf{Inference Efficiency}: Crucially, the inference latency of \textbf{RoDPO} is nearly identical to the baseline ($0.45$s vs. $0.42$s per batch). Since the reference model is discarded during deployment and the Sparse MoE only activates a subset of experts, \textbf{RoDPO} significantly improves performance (NDCG@5: $0.0521$ vs. $0.0495$) with almost no additional deployment cost.
\begin{figure}[h]
  \centering
  \includegraphics[width=\linewidth]{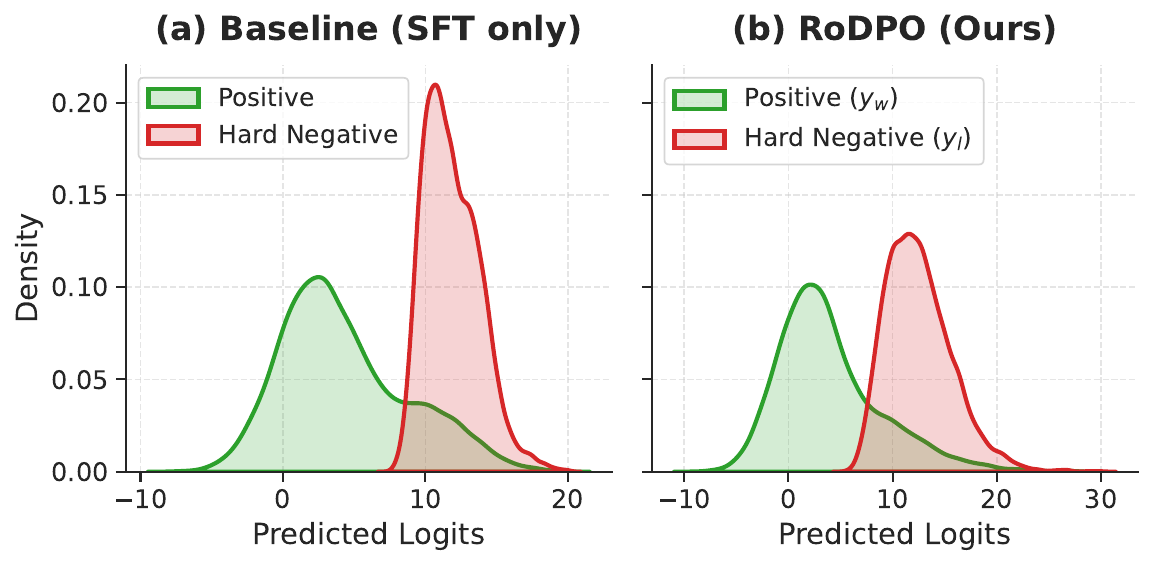}
  \caption{
    Kernel Density Estimation (KDE) of predicted logit distributions for positive items ($y_w$) and hard negative items ($y_l$) on the Amazon Toys and Games dataset.
    }
  \label{fig:logits_dist}
\end{figure}

\subsection{Visualization of Preference Alignment}To intuitively assess preference alignment, we visualize the predicted logit distributions for positive ($y_w$) and hard negative ($y_l$) items in Figure~\ref{fig:logits_dist}.

\textbf{Analysis of Overconfidence in Baseline:} As shown in Figure~\ref{fig:logits_dist}(a), the SFT baseline exhibits a sharp, high-density peak for hard negatives (red) with higher logits than the positives (Green). This confirms our hypothesis regarding the \textit{False Negative} problem: the model is excessively confident in ranking unobserved items higher than the ground truth, suggesting severe overfitting on noisy negative signals.

\textbf{Effectiveness of RoDPO:} In contrast, Figure~\ref{fig:logits_dist}(b) shows that RoDPO effectively \textbf{flattens the negative distribution} and reduces the peak density. The more dispersed negative logits indicate that our stochastic sampling mitigates rigid overconfidence in specific false negatives. By ``softening'' these competitive negative signals, RoDPO creates a more robust probabilistic margin, allowing true preferences to emerge more effectively in the top-$K$ rankings.

\subsection{Case Study}

\begin{figure}[t]
  \centering
  \includegraphics[width=\linewidth]{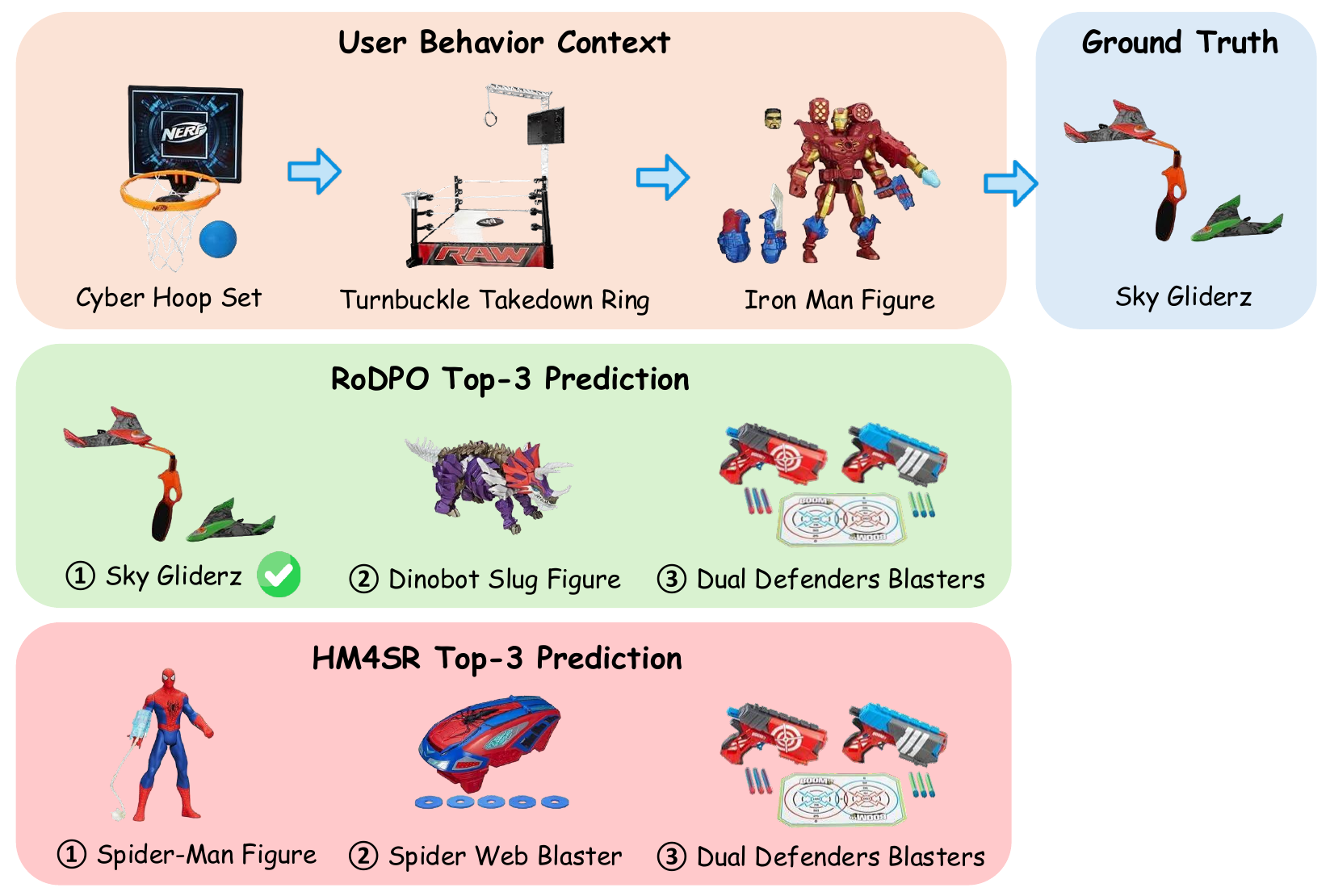}
  \caption{
    A qualitative case study on the Amazon Toys and Games dataset. 
    }
  \label{fig:case_study}
\end{figure}

Figure \ref{fig:case_study} illustrates a representative case. The user shows diverse interests in active sports (\textit{Cyber Hoop}) and action figures (\textit{Iron Man}). However, the baseline \textbf{HM4SR} falls into a \textit{similarity trap}, over-emphasizing the recent \textit{Iron Man} interaction to recommend a semantically similar \textit{Spider-Man} figure. 
In contrast, \textbf{RoDPO} effectively suppresses this hard negative to identify the ground truth \textit{Sky Gliderz}. By prioritizing preference ranking, RoDPO captures latent ``projectile/flight'' intents from earlier history (e.g., \textit{Cyber Hoop}), validating that our Stochastic Top-K Sampling effectively distinguishes true user interests from misleading semantic matches.

\section{Conclusion}

In this paper, we explored the potential of aligning multimodal sequential recommendation models with user preferences via Direct Preference Optimization (DPO). We identify the ``False Negative'' dilemma as a critical barrier when transferring NLP-based alignment strategies to RecSys. To address this, we propose \textbf{RoDPO}, which incorporates a \textit{Stochastic Top-K Negative Sampling} strategy and a \textit{Sparse Mixture-of-Experts} backbone to balance optimization stability with model capacity. Empirical results on three real-world datasets demonstrate that RoDPO significantly outperforms state-of-the-art baselines while maintaining inference efficiency. Our findings provide a robust methodology for aligning recommendation models with user preferences, paving the way for future research on larger-scale foundation models and diverse feedback signals.

\section{Limitations}
While RoDPO achieves robust performance, there are limitations. First, our evaluation is primarily conducted on e-commerce datasets (Amazon), and the generalization to other multimodal domains (e.g., short-video recommendation) remains to be verified. Second, although the Sparse MoE reduces inference latency, the training cost is still higher than lightweight ID-based models due to the additional reference model forward pass. Future work will address efficiency in larger-scale foundation model adaptation.

\bibliography{custom}

\clearpage
\appendix
\section{Appendix}
\label{sec:appendix}

\subsection{Statistics of Datasets}
\label{sec:dataset_statistics}
We present the statistics of processed \textbf{Toys and games}, \textbf{Beauty} and \textbf{Home and Kitchen} datasets in Table~\ref{tab:dataset_statistics}.

\begin{table}[h]
\centering
\resizebox{1.0\columnwidth}{!}{%
\begin{tabular}{lccc}
\toprule
\textbf{Datasets} & \makecell{\textbf{Toys and} \\ \textbf{Games}} & \makecell{\textbf{Beauty}} & \makecell{\textbf{Home and} \\ \textbf{Kitchen}} \\
\midrule
\# Users          & 19,412    & 22,363  & 66,520  \\
\# Items          & 11,924    & 12,101  & 28,238  \\
\# Actions        & 167,597  & 198,502 & 551,682 \\
Avg. Actions/User & 8.63       & 8.88    & 8.29    \\
Avg. Actions/Item & 14.06      & 16.40   & 19.54   \\
\# Sparsity       & 99.93\%  & 99.93\% & 99.97\% \\
\bottomrule
\end{tabular}%
}
\caption{Statistics of processed datasets.}
\label{tab:dataset_statistics}
\vspace{-17pt}
\end{table}

\subsection{Baseline Methods}
\label{sec:baselines}
We compare our proposed method with representative baselines from three categories:
(1) \textbf{Traditional Non-Time-Aware Models}: \textbf{GRU4Rec} \cite{GRU4Rec} and \textbf{SASRec} \cite{SASRec}, which adopt GRUs and self-attention mechanisms respectively; and \textbf{LRURec} \cite{LRURec}, utilizing linear recurrent units for long sequence encoding.
(2) \textbf{Traditional Time-Aware Models}: \textbf{TiSASRec} \cite{TiSASRec}, designing time interval-aware self-attention; \textbf{FEARec} \cite{FEARec}, analyzing frequency-level sequences with a ramp structure; and \textbf{TiCoSeRec} \cite{TiCoSeRec}, which applies contrastive learning on time-augmented sequences.
(3) \textbf{Multimodal Models}: \textbf{NOVA} \cite{NOVA} and \textbf{DIF-SR} \cite{DIF-SR}, employing non-invasive attention and decoupled fusion respectively; \textbf{UniSRec} \cite{UniSRec}, transferring textual semantics via MoE; \textbf{MISSRec} \cite{MISSRec}, featuring an Interest Discovery Module; \textbf{M3SRec} \cite{M3SRec}, utilizing cross-modal MoE transformers; \textbf{IISAN} \cite{IISAN}, leveraging parameter-efficient fine-tuning for intra- and inter-modal adaption; \textbf{TedRec} \cite{TedRec}, performing semantic fusion via Fast Fourier Transform; and \textbf{HM4SR} \cite{HM4SR}, a state-of-the-art baseline employing a hierarchical time-aware MoE framework.

\subsection{Evaluation Metrics}
\label{sec:metrics_detail}
To evaluate the performance of our model, we employ two widely used ranking metrics: Normalized Discounted Cumulative Gain (\textbf{NDCG@K}) and Mean Reciprocal Rank (\textbf{MRR@K}). The formal definitions are as follows:

\begin{itemize}
    \item \textbf{NDCG@K}: This metric measures the ranking quality by considering the position of the ground-truth item in the top-$K$ list. It is defined as:
    \begin{equation}
        \text{NDCG@}K = \frac{1}{|U|} \sum_{u \in U} \frac{1}{\log_2(\text{rank}_{u} + 1)}
    \end{equation}
    where $\text{rank}_{u}$ is the rank position of the target item for user $u$, provided that $\text{rank}_u \leq K$; otherwise, the score is 0.

    \item \textbf{MRR@K}: This metric assesses the average of the reciprocal ranks of the target items. It is calculated as:
    \begin{equation}
        \text{MRR@}K = \frac{1}{|U|} \sum_{u \in U} \frac{1}{\text{rank}_{u}}
    \end{equation}
    Similarly, if the rank position $\text{rank}_u > K$, the reciprocal rank is set to 0.
\end{itemize}
For both metrics, higher values indicate better ranking performance.

\end{document}